\documentclass[preprint]{aastex}
\pdfoutput=1

\usepackage{amsmath}
\usepackage{subfigure}
\usepackage{natbib}
\usepackage{graphicx}
\usepackage{ccaption}

\begin{document}

\title{VLA Observations of DG Tau's Radio Jet: A highly collimated thermal outflow}

\author{C. Lynch$^1$, R. L. Mutel$^1$, M. G\"udel$^2$, T. Ray$^3$, S. L. Skinner$^4$, P. C. Schneider$^5$ \&  K. G. Gayley$^1$}
\affil{$^1$Department of Physics and Astronomy, University of Iowa, Iowa City, Iowa 52240, USA}
\affil{$^2$Department of Astrophysics, University of Vienna, Vienna, AT}
\affil{$^3$Dublin Institute for Advanced Studies, Astronomy and Astrophysics Section, 31 Fitzwilliam Place, Dublin 2, Ireland}
\affil{$^4$Center for Astrophysics and Space Astronomy, University of Colorado, Boulder, Colorado 80309, USA}
\affil{$^5$Hamburger Sternwarte, Gojenbergsweg 112, 21029 Hamburg, Germany}
\begin{abstract}

The active young protostar DG Tau has an extended jet that has been well studied at radio, optical, and X-ray wavelengths. We report sensitive new  VLA full-polarization observations of the core and jet between 5~GHz and 8~GHz. Our high angular resolution observation at 8~GHz clearly shows an unpolarized inner jet with a size 42 AU (0.35\arcsec) extending along a position angle similar to the optical-X ray outer jet.  Using our nearly coeval 2012 VLA observations, we find a spectral index  $\alpha = +0.46\pm0.05$, which combined with the lack of polarization, is consistent with bremsstrahlung (free-free) emission, with no evidence for a non-thermal coronal component. By identifying  the end of the radio jet as the optical depth unity surface, and calculating the resulting emission measure, we find our radio results are in agreement with previous optical line studies of electron density and consequent mass-loss rate. We also detect a weak radio knot at 5~GHz located 7\arcsec\ from the base of the jet, coincident with the inner radio knot detected by \citet{Rodriguez:2012} in 2009 but at lower surface brightness. We interpret this as due to expansion of post-shock ionized gas in the three years between observations.

\end{abstract}

\section{Introduction}

The evolution of a young stellar object (YSO) involves not only mass accretion, through a circumstellar molecular disk, but also the loss of angular momentum and mass flux through a narrowly collimated jet,  thought to be launched close to the YSO.   Magnetic fields are suspected of playing an important role in both the launching and collimation of these jets \citep{Pudritz:2012, Cai:2008}. However, the processes through which these fields act are not well understood \citep[e.g.][]{Carrasco:2010}.

Classical T Tauri stars (CTTS) are YSOs which have evacuated most of their remnant circumstellar envelope. This allows an unobscured line of sight to the central 100 AUs   of their jet and measurements of forbidden line emission are possible \citep{Dougados:2000}. Using the spectroscopic diagnostic technique described by \citet{Bacciotti:1999}, the forbidden-line emission can provide constraints for the electron density, total hydrogen density, ionization fraction, and the average excitation temperature of the gas in these jets. Consequently, probing the inner wind structure of CTTS is crucial in understanding the mechanisms by which these YSO jets are launched and collimated. 

Some CTTS have associated parsec-scale jets, with Herbig-Haro (HH) objects forming where the fast streams of material collide with the slower material along the jet \citep{McGroarty:2007, McGroarty:2004}.  These jets are dynamical, evolving on the timescales of a few years, and contain knots which have proper motions in the range of a few 100 km $\text{s}^{-1}$. In addition, about 10 have been discovered to have X-ray emission.  These X-rays most likely trace the fastest shocks in the jets of YSOs. If the material in the YSO jet is heated to X-ray emitting temperatures by shocks, shock velocities around 500$\ \text{km}\ \text{s}^{-1}$ required \citep{Schneider:2011}.

A large number of the compact jets associated with YSOs have been detected at radio wavelengths. The dominant radio emission mechanism is thought to be thermal bremsstrahlung from the shock-heated gas. However, evidence for non-thermal emission from several YSOs has also been discovered. Polarization has been associated with several YSOs, including objects in the $\rho$ Ophiuchi molecular cloud \citep{Andre:1988, White:1992, Andre:1992}, the Taurus-Auriga molecular cloud \citep{Phillips:1993, Skinner:1993, Feigelson:1994, Ray:1997}, the R Coronae Australis region \citep{Choi:2008}, the Orion region \citep{Zapata:2004}, and HH 7-11 \citep{Rodriguez:1999}. Furthermore, a few YSOs have been associated with linearly polarized radio emission, these include HH 80-81 \citep{Carrasco:2010}, HD 283447 \citep{Phillips:1996}, and the Orion Streamers \citep{Yusef-Zadeh:1990}. Additional characteristics of non-thermal emission, including strong variation in flux density on timescales of hours to days, a negative spectral index, and VLBI measurements of high brightness temperatures ($T_B\gg 10^7$ K), have also been found \citep{Curiel:1993, Hughes:1997, Andre:1992, Wilner:1999, Rodriguez:2005}. A suggested source of this non-thermal emission is the gyrosynchrotron mechanism \citep{Andre:1996}. 

Non-thermal emission is almost always  associated with the more evolved weak-lined T Tauri stars (WTTS). This is consistent with the idea that the non-thermal coronal emission is revealed only after the optically thick mass outflows have evolved away \citep{Eisloffel:2000}. However, this idea may be too simple: Apparently non-thermal emission has recently been reported from several  less-evolved YSO's that have infrared evidence for a disk \citep{Osten:2009}.
   
This paper focuses on DG Tau, a highly active CTTS driving a well studied energetic bipolar jet. Located in the Taurus Molecular Cloud (estimated distance 140 pc; \citet{Torres:2009}), DG Tau is of spectral type K5-M0, with a mass of 0.67 $M_{\odot}$, a luminosity of 1.7 $L_{\odot}$ and an estimated age of 3 $\times$ $10^5$ yr \citep{Kitamura:1996, Gudel:2007}. It was one of the first T Tauri stars to be associated with an optical jet \citep{Mundt:1983} and has been studied extensively with adaptive optics, interferometry and the $Hubble$ $Space$ $Telescope$. The ``micro-jet", associated with the HH 158 knot, extends out to 12''\citep[e.g.][]{Eisloffel:1998} at a position angle of $223^{\circ}$ \citep{Lavalley:1997}.  \citet{Eisloffel:1998} calculated the proper motion of 4 knots observed in the HH 158 jet located at distances between  $2.5\arcsec$-$10.0\arcsec$ ( 350-1400 au), velocities around 150$\ \text{km}\ \text{s}^{-1}$. Furthermore, \citet{Dougados:2000} calculated the proper motions of knots located within $4.0\arcsec$ from the source of the HH 158 jet; they determine velocities of $\sim$200$\ \text{km}\ \text{s}^{-1}$. 

The jet of DG Tau has an onion-like structure within 500 AU of the star, where faster, highly collimated gas is nested in wider slower material, with maximum bulk gas speeds reaching 500$\ \text{km}\ \text{s}^{-1}$ \citep{Bacciotti:2000, Lavalley-Fouquet:2000}. This velocity structure is expected if the jet material is launched from a range of disk radii \citep{Agra-Amboage:2011}. Moreover, for the slower jet material in the jet the velocity between the two sides of the jet is between 6-15$\ \text{km}\ \text{s}^{-1}$; this velocity shift may indicate that the jet is rotating \citep{Bacciotti:2002b}. Using [Fe II] observations and averaging over the central $1.0\arcsec$ ,  the mass-loss rate from the high velocity gas is determined to be (1.6$\ \pm\ $0.8)$\times$$10^{-8}$$M_{\odot}$$\ \text{yr}^{-1}$ and from the medium velocity gas (1.7$\ \pm\ $0.7)$\times$$10^{-8}$$M_{\odot}$$\ \text{yr}^{-1}$, giving a total mass-loss rate for the velocity range of 50 to 300$\ km\ \text{s}^{-1}$ of (3.3$\ \pm\ $1.1)$\times$$10^{-8}$$M_{\odot}$\ $\text{yr}^{-1}$. However, this value is a lower limit to the mass-loss rate from the atomic component of the jet since the [Fe II] emission does not probe the whole range of velocities seen at optical wavelengths \citep{Agra-Amboage:2011}. The mass-loss rate of DG Tau has also been estimated using [OI]$\lambda$6300, giving a rate of 3$\times$$10^{-7}$$M_{\odot}$\ $\text{yr}^{-1}$ \citep{Hartigan:1995}.  The most recent estimate for the mass loss through the jet atomic component is  \citet{Maurri:2012},  with $\dot{M}\ =  (1.2\ \pm\ 0.4) \times\ 10^{-8}\ M_\odot\ yr^{-1}$. Near-infrared evidence for a counter-jet has been reported by \citet{Pyo:2003a}. The redshifted emission appears suddenly at -$0.7\arcsec$ which suggests that the inner part of the counter-jet is hidden behind an optically thick circumstellar disk.  

DG Tau shows strong millimeter continuum emission thought to arise from a compact dust disk around the star.  Assuming a  dust opacity coefficient $\kappa_{\nu}$=0.02-0.05 $\text{cm}^2$ $\text{g}^{-1}$ at 147 GHz, the spectrum is consistent with thermal emission from a disk having a radius of about 110 AU and a mass 0.01-0.06 $M_{\odot}$  \citep{Kitamura:1996}. At larger scales, $^{13}\text{CO}$ observations have revealed a gas disk with radius of 2800 au, oriented with its major axis perpendicular to the jet of DG Tau \citep{Kitamura:1996a}. 

The X-ray jet of DG Tau was first discovered by \citet{Gudel:2005} using a $Chandra$ ACIS-S observation, which showed very faint soft emission along the jet out to a distance of about $5.0\arcsec$ to the SW with position angle $225^{\circ}$.  In addition, \citet{Gudel:2005} found that DG Tau reveals a new type of X-ray spectrum that includes two emission components with vastly different absorbing column densities: A weakly attenuated soft spectral component associated with a plasma with electron temperatures of no more than a few MK and a strongly absorbed hard spectral component associated with a hot plasma of several tens of MK.The soft X-ray component is thought to originate at the base of the jet where the first shocks form, while the hard component is attributed to emission arising from a magnetospheric corona. Moreover, \citet{Schneider:2008} demonstrated that the soft and hard X-ray components have a separation of $\sim$ $0.2\arcsec$ and that the soft X-ray emission is coincident with emission from optical lines indicating that this X-ray component is indeed from the jet. 

There are few previous radio observations of DG Tau. It was first detected in a VLA survey of the Taurus-Auriga region at 5 GHz \citep{Cohen:1982}. Further observations  at 15 GHz and 1.5 GHz \citep{Cohen:1986} found that the structure is elongated. \citet{Cohen:1986} suggested that the radio structure and spectrum is consistent with free-free emission from ionized gas in an outflowing jet. \citet{Rodriguez:2012} reported radio knots in the extended jet at $0.42\arcsec$ and $6.98\arcsec$ respectively. They determine that these radio components were coincident with previously detected optical knots. 

In this paper we present the results of a Jansky Very Large Array \footnote{The VLA is operated by the National Radio Astronomy Observatory, which is a facility of the National Science Foundation operated under cooperative agreement by Associated Universities, Inc.} (VLA) multi-frequency campaign of DG Tau.  The goals of these observations were to determine the morphology of the inner jet,  search for possible non-thermal coronal radio emission, and investigate the nature of the radio knots at large distance from the star.  
\section{VLA Observations}
\begin{deluxetable}{ccccccccc}
\tablecolumns{9}
\tablecaption{VLA Results}
\tablehead
{
\colhead{Epoch}& 
\colhead{Array} & 
\colhead{Freq} & 
\colhead{BW} & 
\colhead{Time} & 
\colhead{Total flux} & 
\colhead{Peak flux} & 
\colhead{Sky RMS} &
\colhead{$\Theta_{FWHM}$} \\

\colhead{} &
\colhead{} &
\colhead{(GHz)} &
\colhead{(GHz)} &
\colhead{(min)} &
\colhead{(mJy)} &
\colhead{(mJy/beam)} &
\colhead{($\mu$Jy)} &
\colhead{(arcsec)}
}
\startdata
2000.83 & A & 43.3 & 0.1     &  240 & 3.9 &  0.6 & 100 &   0.05$\times$0.04\\
2011.46 & A &  8.5 & 0.256 &  198 &  1.1 &  0.4 & 10.0 &  0.38$\times$0.26 \\
2012.22 & C & 5.4  & 2.0    &     29 & 1.05 & 1.02 & 6.2 &  4.3$\times$4.0  \\
2012.29 & C &  8.5 & 2.0    &     30 & 1.29  & 1.16 & 8.0 &  2.7$\times$2.4     \\
\enddata
\label{table:obs}
\end{deluxetable}

The radio campaign comprised three observing epochs between June 2011 and April 2012.  The June 18, 2011 observation used two 128 MHz bands centered on 8.33 GHz and 8.46 GHz  in A configuration. In 2012 we conducted two C-array observations using the newly-available 2~GHz bandwidth capability of the VLA. The March 22, 2012 observation spanned the frequency range of 4.5 GHz to 6.5 GHz, while the April 15, 2012 observation  spanned 7.9 GHz to 9.9 GHz. The details of these observations are listed in Table \ref{table:obs}. 

For all three observations both the receiver bandpass correction and the absolute flux density scale was set using the amplitude calibrator 3C48. The phase calibration used the angularly nearby phase calibrator source J0403+2600.  In addition,  we made short observations of both 3C84 and 3C138 in our June 2011 observation.  These sources were used to solve for the antenna polarization leakage terms and linear polarization position angle. Data reduction and imaging was carried out using Common Astronomy Software Application (CASA) package distributed by the National Radio Astronomy Observatory (NRAO). Additionally, we used the Astronomical Imaging Processing System (AIPS) to calibrate and map an archival 4-hour A-array VLA observation of DG Tau from November 1, 2000 at 43~GHz. 

\section{Results}
\subsection{Imaging the DG Tau Jet}

\begin{figure}[h!]
\epsscale{0.8}
\plotone{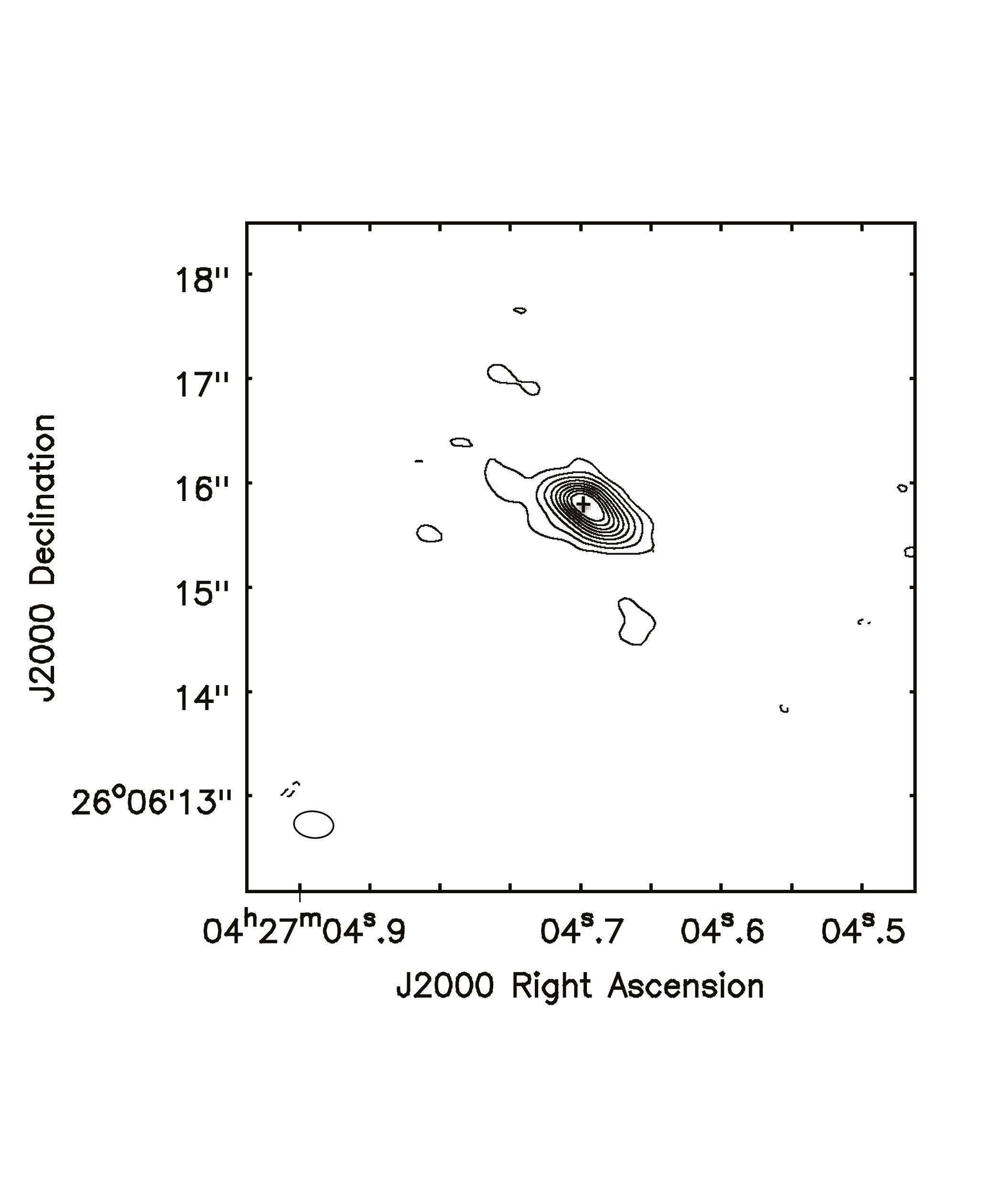}
\caption{VLA 8~GHz naturally-weighted contour map of DG Tau's inner jet at epoch  2011.46. The contours are -3, 3, 6, 10, 15, 20, 25, 30, 35, and 40 $\times$ 10 $\mu$Jy beam$^{-1}$, the RMS noise level of the image. The restoring beam is shown in the bottom left corner with dimensions given in Table$\  \ref{table:obs}$.}
\label{fig:X2011}
\end{figure}

After flagging bad visibilities and applying the normal amplitude and phase calibration, clean maps in total intensity (Stokes $I$, see Figure \ref{fig:X2011}), circularly polarized intensity (Stokes $V$) and linear polarized intensity were produced using the standard CASA data calibration procedures. The peak and integrated flux density of both the total and polarized intensity was obtained using the CASA task IMFIT (see Table \ref{table:obs}).  We did not detect either circular or linear polarization, with upper limits  0.03 mJy beam$^{-1}$ (3\%) for circular and 0.02 mJy beam$^{-1}$ (2\%) for linear polarization. 
Both the rising spectral index and lack of polarization suggest that any non-thermal coronal component is absent or very weak. 

The A-array 8~GHz  total intensity contour map at epoch 2011.46 is shown in Figure \ref{fig:X2011}.  In this high-angular resolution image, the emission extends  approximately 0.35\arcsec\ SW in approximately same direction as the optical jet. Both the asymmetry and elongation of the emission suggest that the source is the collimated jet mapped by optical observations \citep{Bacciotti:2000, Agra-Amboage:2011} which have resolved the inner jet to within $0.1\arcsec$ of the central source and determined that collimation must occur within this region. 
\begin{figure}[h!]
\epsscale{0.99}
\plottwo{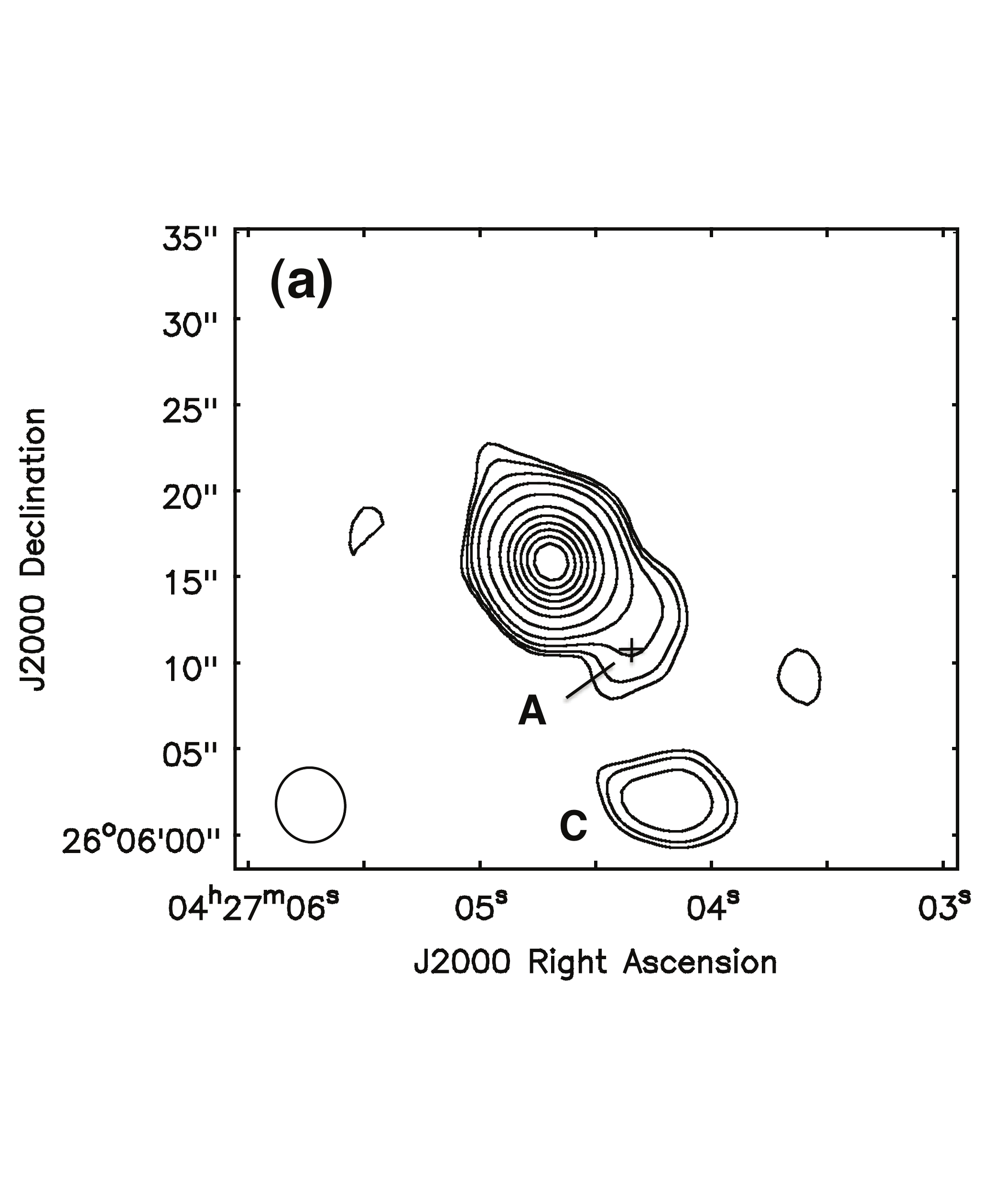}{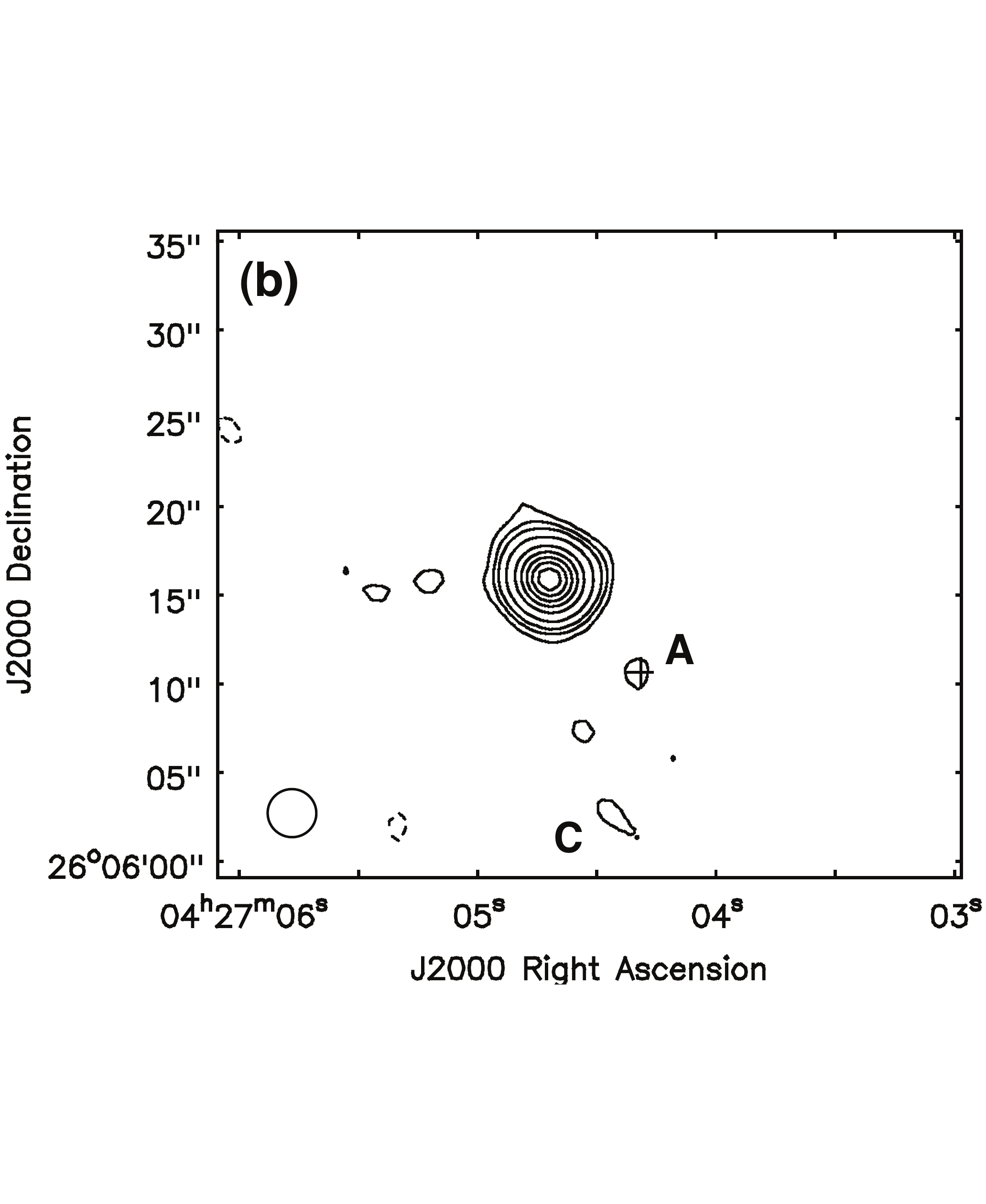}
\caption{VLA contour maps of DG Tau during the two spring 2012 observations. The 8.5 GHz map at epoch 2012.29 is shown on the right, while the 4.5-6.5~GHz emission (epoch 2012.22) map is on the left.  Both maps were made used natural weighting with contour levels -3, 3, 4, 6, 8, 10, 12, 15,  20, 40, and 60 $\times$ 8.0 and 6.2 $\mu$Jy beam$^{-1}$, the RMS values of the respective maps. The restoring beam for each map is given in the bottom left corner with the dimensions given in Table~\ref{table:obs}. The cross  indicates the location of knot A  detected by\protect \citet{Rodriguez:2012} in 2009 radio observations. Feature C, located $\sim$$20.0\arcsec$ SW of the inner centriod, is visible in both 2012 maps and on the 2009 map of \protect \citet{Rodriguez:2012}, but is displaced from the optical jet.} 
\label{fig:CXBandNat}
\end{figure}

The 2012 observations were taken in C-array, with much lower angular resolution but higher sensitivity to larger, low-surface brightness components.  Contour maps of the 5~GHz (epoch 2012.22) and 8~GHz (epoch 2012.29) observations are shown in Figure \ref{fig:CXBandNat}. The location of the central source for the 2012.22 (5.5 GHz) observation is coincident with  the 2012.29 (8.5 GHz) central source within the centroid uncertainty ($\pm\ $$0.01\arcsec$). Note that the integrated flux density at 8~GHz apparently increased 17\% from epoch  2011.46 to 2012.29, but it is possible that the increase results from faint extended structure over-resolved by the smaller beam at epoch 2011.46 (cf. Table 1). 
The 5.5~GHz map shows a component (labeled A) extending $\sim7\arcsec$ SW whose centroid is coincident with a much weaker feature seen in the 
8.5~GHz map.  This component is  nearly coincident with the knot component reported by \citet{Rodriguez:2012} but at a much lower flux density; we discuss this in more detail in  section 4.3. 

There is also a more distant feature in the 5~GHz map (labeled C) with a flux density $S\sim100$ $\mu$Jy about 14\arcsec\ from the stellar position, but well-displaced from the jet axis. This feature is also weakly seen in the Rodriguez et al. map, and is certainly real. We do not know if it is associated with the DG Tau jet, although it is unlikely that an unrelated background source this strong would be located this close to the jet. We speculate that the feature may be a part of an extended bow shock associated with the optical knot seen near 13\arcsec\ \citep{Rodriguez:2012}, but do not discuss it further in this paper.
\subsection{Spectral energy distribution}
\begin{figure}[h!]
\centering
\includegraphics[width=\textwidth]{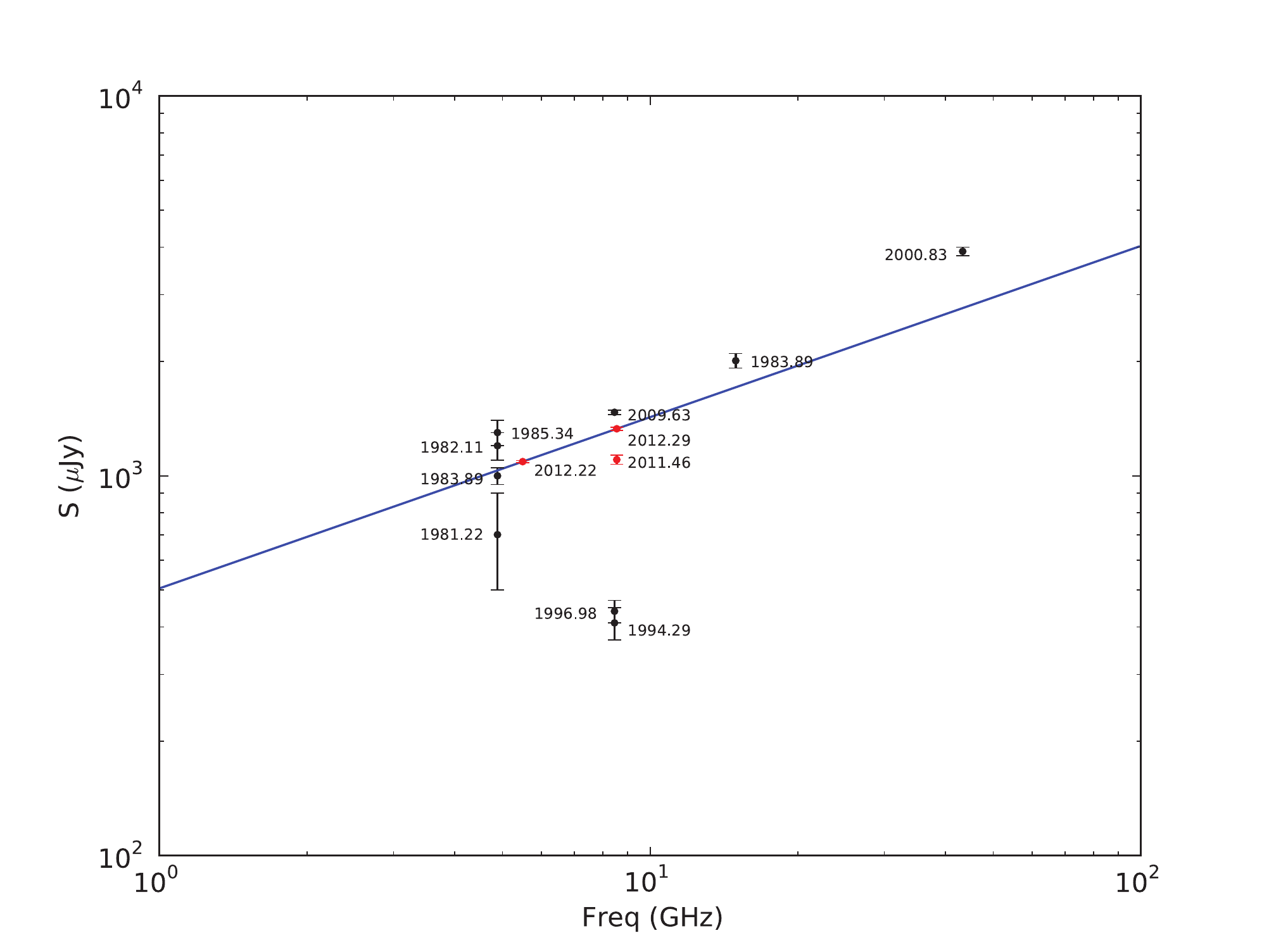}
\caption{Flux density of DG Tau as a function of frequency for observations at epochs 1981.22 --- 1985.34\protect \citep{Cohen:1986},  1994.29, 1996.98, and 2009.63 \protect \citep{Rodriguez:2012},  2000.83  (archival VLA), and 2011.46 --- 2012.29 (this paper). The solid line, a power-law fit to the nearly coeval 2012.22 and 2012.29 points only, has a spectral index $\alpha = 0.46\pm0.05$.}
\label{fig:spec}
\end{figure}

Figure \ref{fig:spec} shows a radio spectrum of DG Tau using our VLA data, previously published radio data \citep{Cohen:1982, Cohen:1986, Rodriguez:2012}, and an archival VLA observation of DG Tau at 43~GHz  (project code AW545, PI David Wilner) that we calibrated and mapped. The spectrum monotonically increases over the frequency range of 4.5 to 43 GHz. However, it is clear from the multiple observations at 8.5 GHz and 4.5 GHz that the flux density is significantly variable over a period of years. Therefore we use only our nearly coeval 2012 observations to fit a power-law to the spectrum, giving a spectral index $\alpha$= 0.46$\ \pm\ $0.05. This spectral index is typical of collimated thermal jets, whose spectral indices are close to +0.6 for constant velocity isothermal outflows, but whose overall spectral index can vary from nearly flat to +2 depending on the physical conditions in the jet, such as velocity gradients and recombination in the flow \citep{Reynolds:1986}. To better constrain the spectral index of the radio emission we need nearly coeval observations over a much larger range of frequencies.

\section{Discussion}
\subsection{Comparison between radio and optically-derived electron densities}

\begin{figure}[t!]
\begin{center}$
\begin{array}{cc}
\includegraphics[width=3.2in]{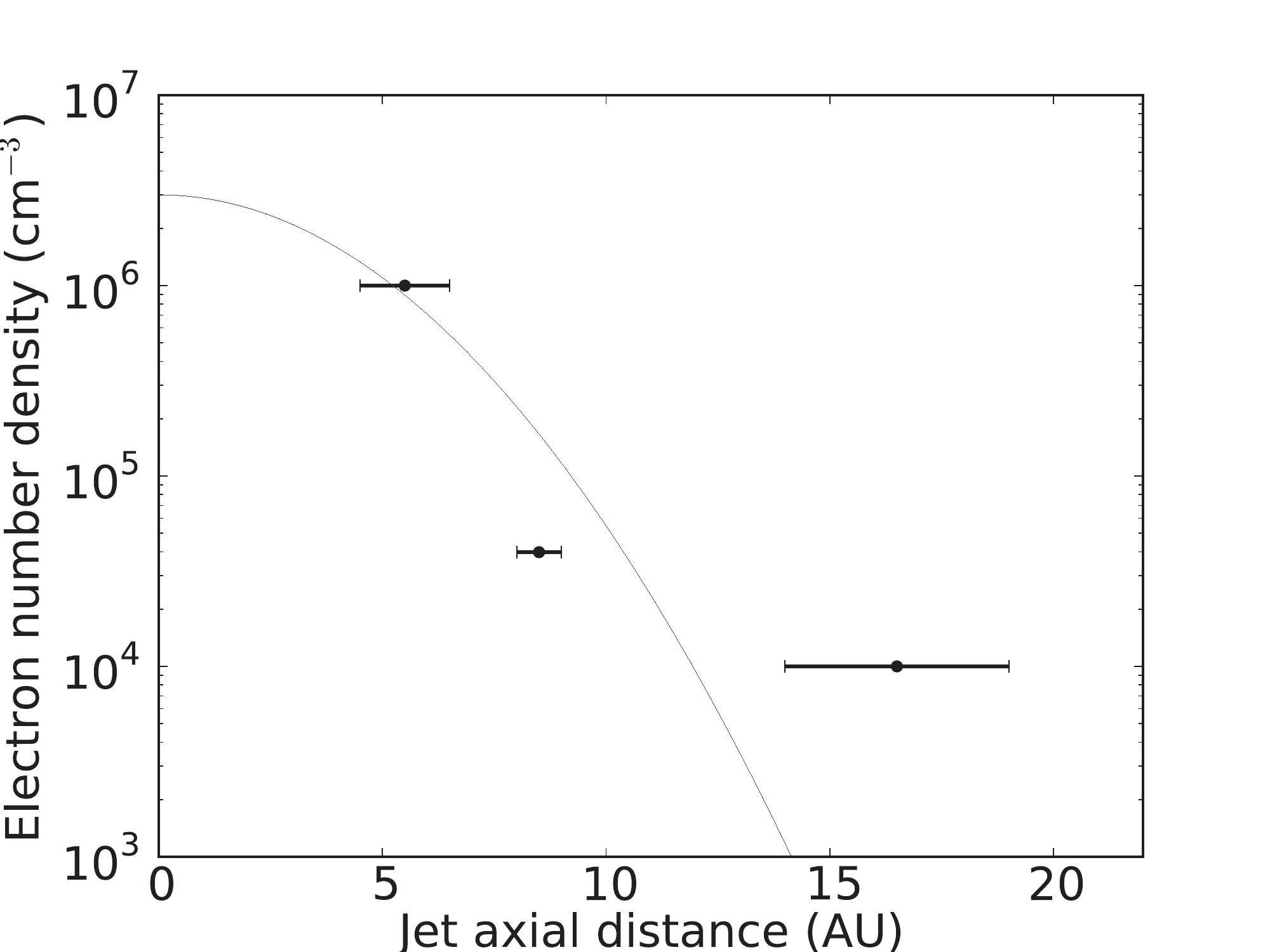} &
\includegraphics[width=3.2in]{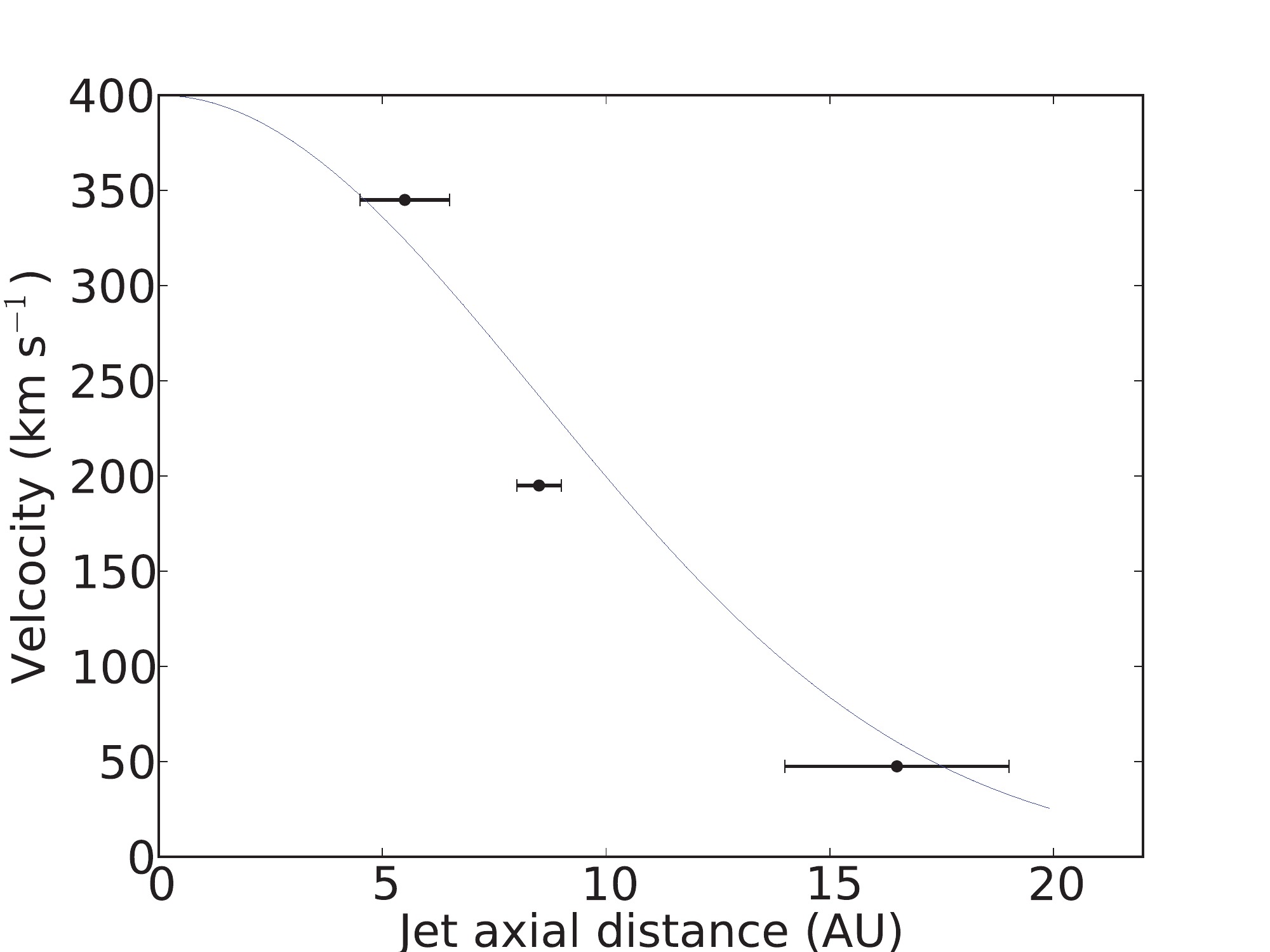}
\end{array}$
\end{center}
\caption{Analytic functions used to model the electron density (left) and velocity (right) profiles across DG Tau's jet at $0.35\arcsec$ from the central source. The points are derived values of electron density and velocity from \protect\citep{Maurri:2012}.}
\label{fig:modelab}
\end{figure}
Free-free emission depends only on the plasma temperature, the observing frequency, and the linear emission measure i.e., the square of the electron density integrated along the line of sight to the observer. If the plasma temperature is known (e.g., from optical line observations), and the optical depth at a given location can be estimated (e.g., from source structure), the emission measure at that location can be estimated. The entire detected radio jet is the optically thick surface. We therefore identify the outermost detectable region of the radio jet with the unity optical depth surface ($\tau =1$). We then can calculate the emission measure  and compare with estimates of electron density and jet width at this location based on optical line ratios. 

Inspection of Figure~1 shows that the radio emission becomes undetectable $\sim0.35\arcsec$ (50 AU projected) from the base of the jet. Assuming a thermal jet, this should correspond to the $\tau=1$ surface, meaning inside $\sim0.35\arcsec$ the jet is fully optically thick while outside this distance the jet is optically thin. The temperature of the jet gas is not well-constrained, but if the jet is launched via a quasi-steady centrifugal MHD disk wind, as suggested by the transverse velocity gradients \citep{Bacciotti:2002}, the gas temperature should be not much higher than the photospheric temperature (4800 K). However, the innermost, high-velocity core is likely shock-heated and may have a temperature near 8000 K, typical to other stellar jets.  The lower-speed, broader flow associated with warm molecular emission is cooler, with a temperature of 2000 K, determined from infrared line ratios \citep{Takami:2004}. In the following we assume a mean temperature across the flow $\overline{T} = 5000$ K. 

The free-free optical depth can be written
\begin{equation}
\tau(\nu,T,EM) = 1.06\ {\left(\frac{\nu}{\rm{GHz}}\right)}^{-2.1} {\left(\frac{T}{10^4\ \rm{K}}\right)}^{-1.35} {\left(\frac{\rm{EM}}{10^{25}\ {\rm{cm}}^{-5}}\right)}
\end{equation}

Assuming $\tau =1$ at $\nu$= 8.5~GHz, the emission measure at 0.35\arcsec\ is
\begin{equation}
EM \equiv \int{{n_e}^2 ds} = 3.3\times10^{26}\ {\rm{cm}}^{-5}
\end{equation}
 
High angular resolution observations of optical line ratios have be used to determine both the electron density and jet width as a function of velocity bin along the jet \citep[e.g.,][]{Coffey:2008, Maurri:2012}. By combining these measurements, we can estimate the axial electron density profile.  Figure~\ref{fig:modelab}(a) shows a simple Gaussian fit to the electron density as a function of axial distance, where the points represent the mean density in each velocity bin and the uncertainties are the FWHM width uncertainty in each bin, using data from \citet{Maurri:2012}. We have combined these measurements to fit a simple analytic function to the electron density as a function of jet axial distance $\rho$,
\begin{equation}
  n_{e}(\rho)=n_{0}\ 
  e^{-\left(\frac{\rho}{\rho_e}\right)^2}
  \end{equation} 
where $n_0  = 2.5\times10^6$ cm$^{-3}$ and $\rho_e$ = 5.5 au. 
Integrating this density profile, we find a linear emission measure  EM = $3.2\times10^{26}$ cm$^{-5}$, in excellent agreement with the radio data. 

\subsection{Mass-loss rate of the ionized component}

We next use the axial electron density derived in the previous subsection to estimate the mass-loss rate of the ionized component of the jet outflow at  $\tau=1$  (50 AU projected distance). Figure~\ref{fig:modelab}(b) shows a Gaussian fit to the FWHM widths of the velocity bins given in \citet{Maurri:2012},
\begin{equation}
  V(\rho)= V_{0}\ 
  e^{-\left(\frac{\rho}{\rho_v}\right)^2}
  \end{equation}
 where $V_{0}$= 400 km $\text{s}^{-1}$, and $\rho_v$= 12 au. Using this velocity profile, along with the density profile determined in the previous sub-section, we determine the mass-flux of the ionized gas at 50 AU projected distance,

\begin{equation}
\dot{M}=2\ \pi\ \overline{m_i} \ \int_0^\infty\  V(\rho)\ n_{e}(\rho)\ \rho\,d\rho. 
\label{eqn-massloss-onion}
\end{equation}
where $\overline{m_i}\sim1.2\ m_p$ is the average ion mass. Using the approximate analytic models shown in Figure~\ref{fig:modelab} for $n_e(\rho)$ and $V(\rho)$, we find mass flux $\dot{M}\sim$ 5 $\times$\ $10^{-8}$\ $M_\odot$\ $yr^{-1}$, with more than half of the mass flux in the high-velocity component within 5 AU of the jet center. Since the high-velocity component is almost completely ionized \citep{Maurri:2012}, this mass-loss estimate should be less than a factor of two smaller than the mass-loss of both the ionized and neutral gas. 

Recent mass loss estimates of the DG Tau jet include that of \citet{Agra-Amboage:2011}, $\dot{M}\ =  (3.3\ \pm\ 1.1) \times\ 10^{-8}\ M_\odot\ yr^{-1}$ and \citet{Maurri:2012}, $\dot{M}\ =  (1.2\ \pm\ 0.4) \times\ 10^{-8}\ M_\odot\ yr^{-1}$. While these are both somewhat lower than the present model prediction, the differences are probably not significant given the poorly-constrained functional forms used for the axial density and velocity profiles. 
 
\subsection{Radio Knots in the Extended Jet}
\begin{figure}[!t]
\begin{center}$
\begin{array}{cc}
\includegraphics[width=2.9in]{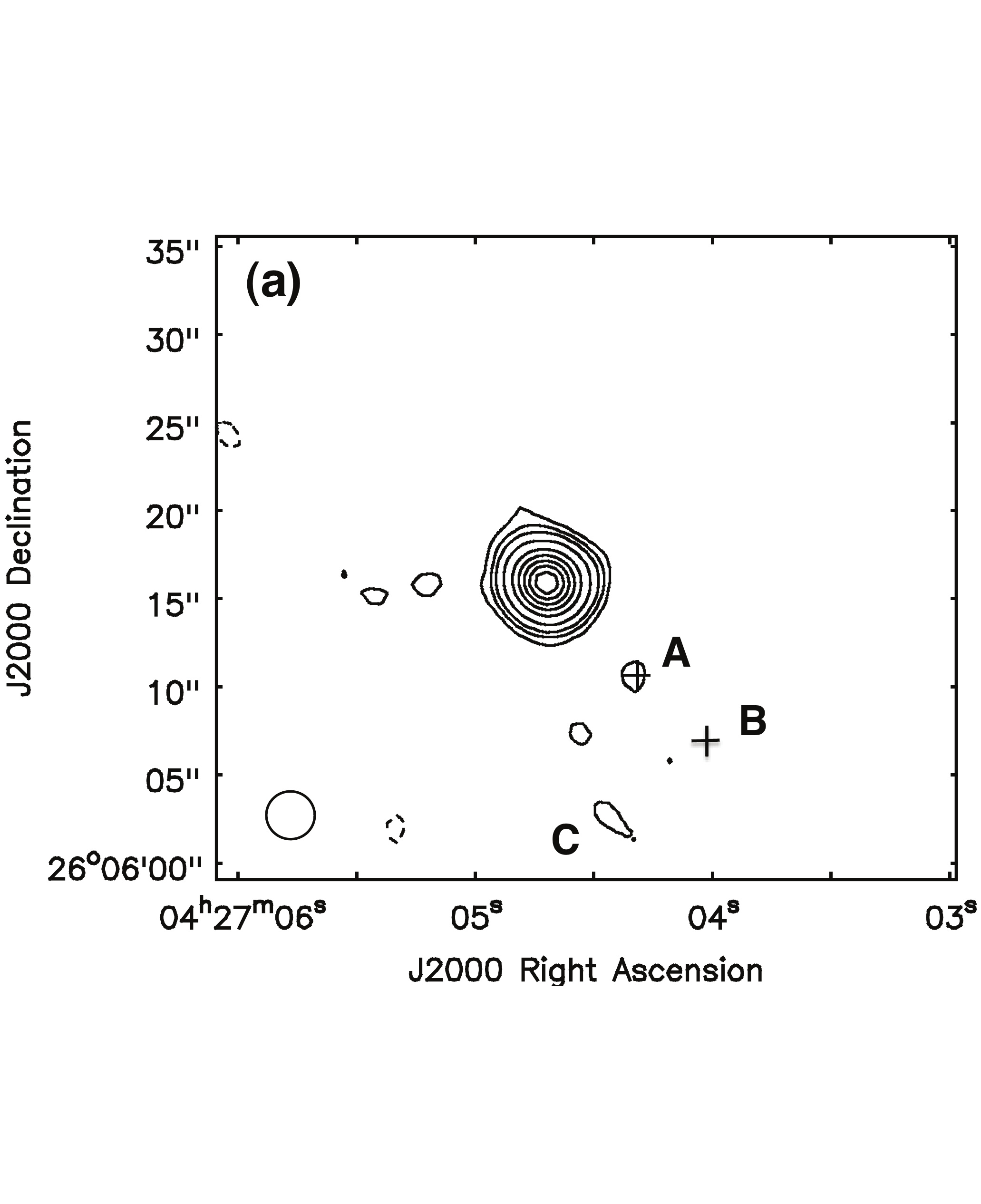} &
\includegraphics[width=2.9in]{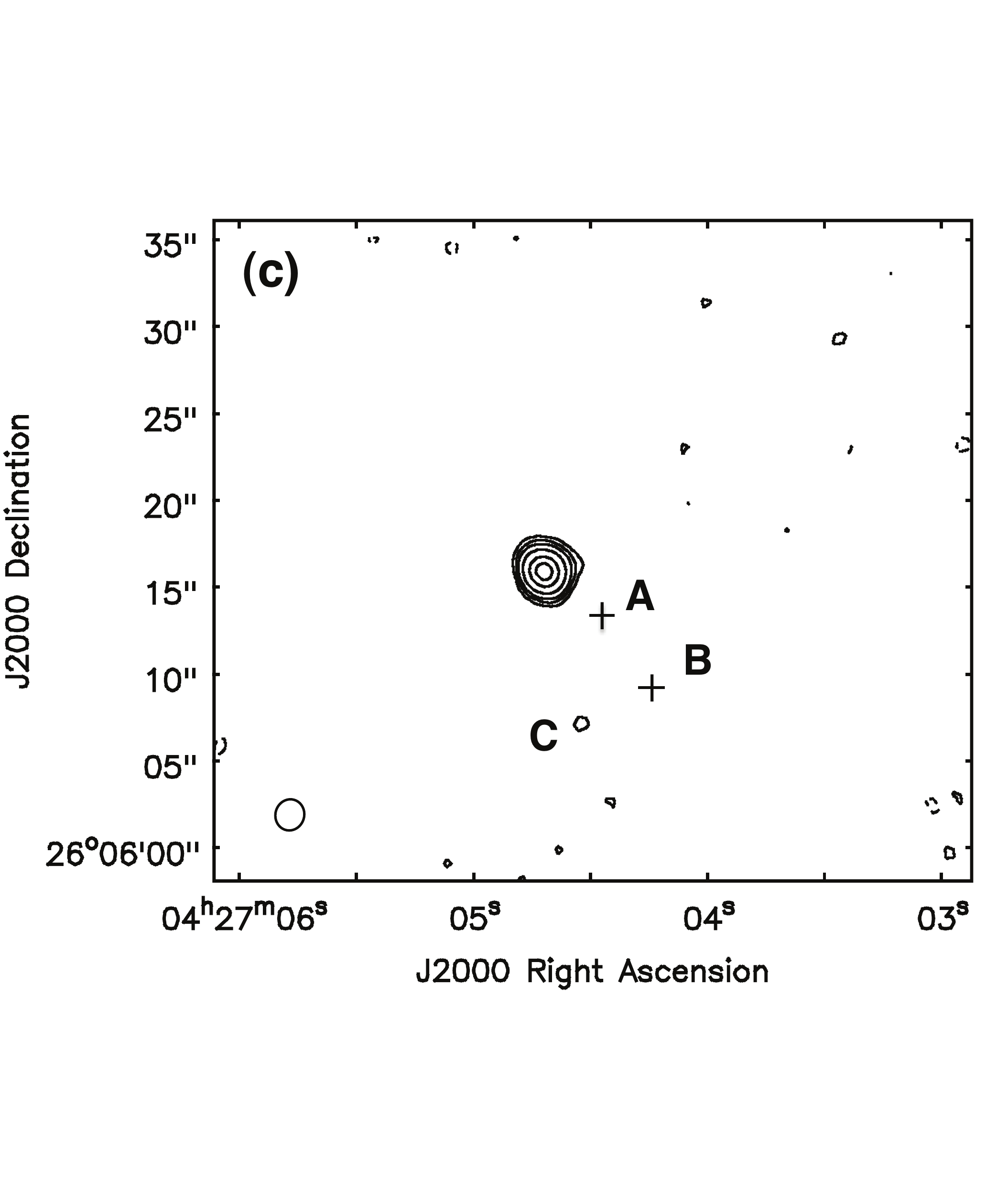} \\
\includegraphics[width=2.9in]{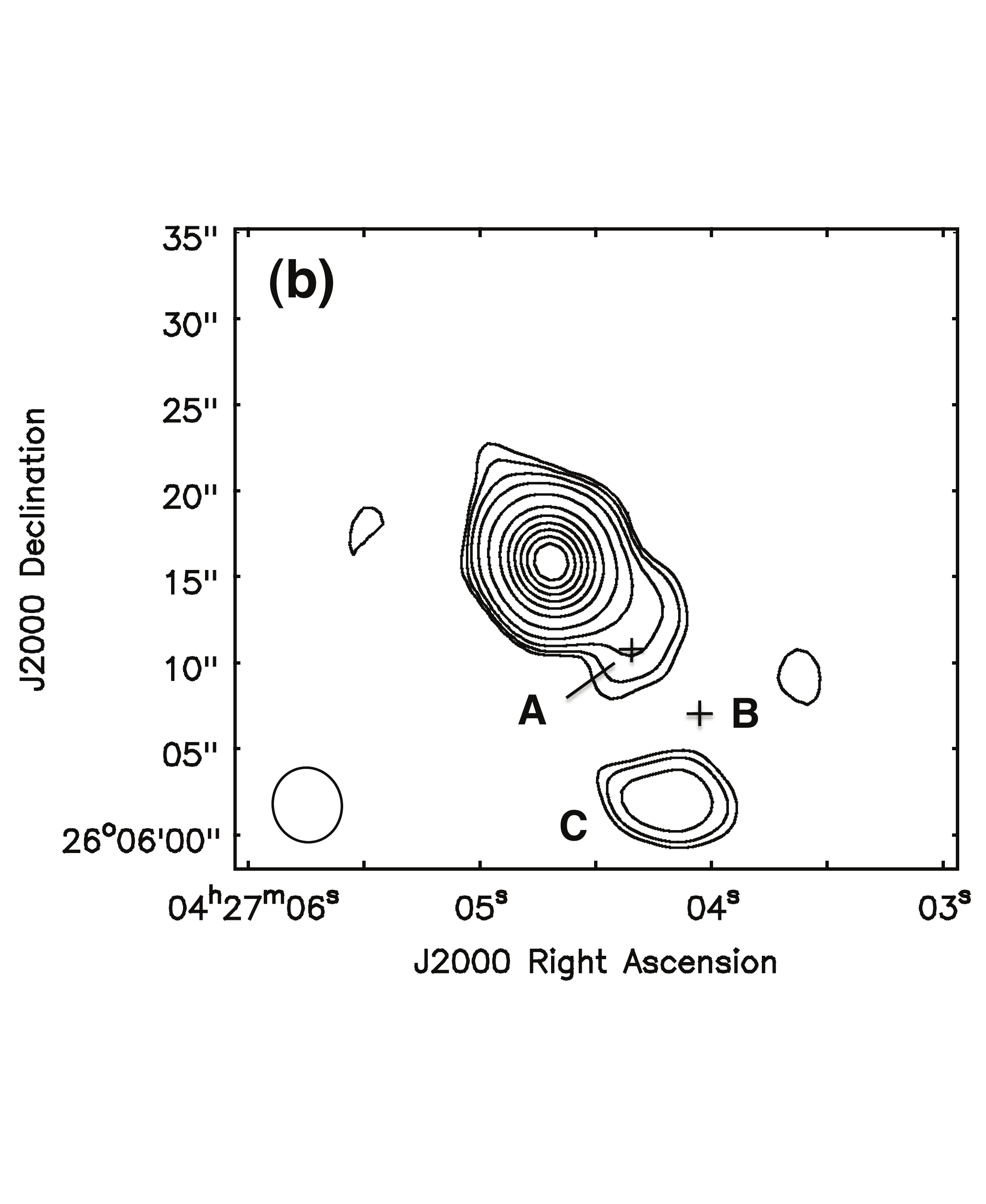} &
\includegraphics[width=2.9in]{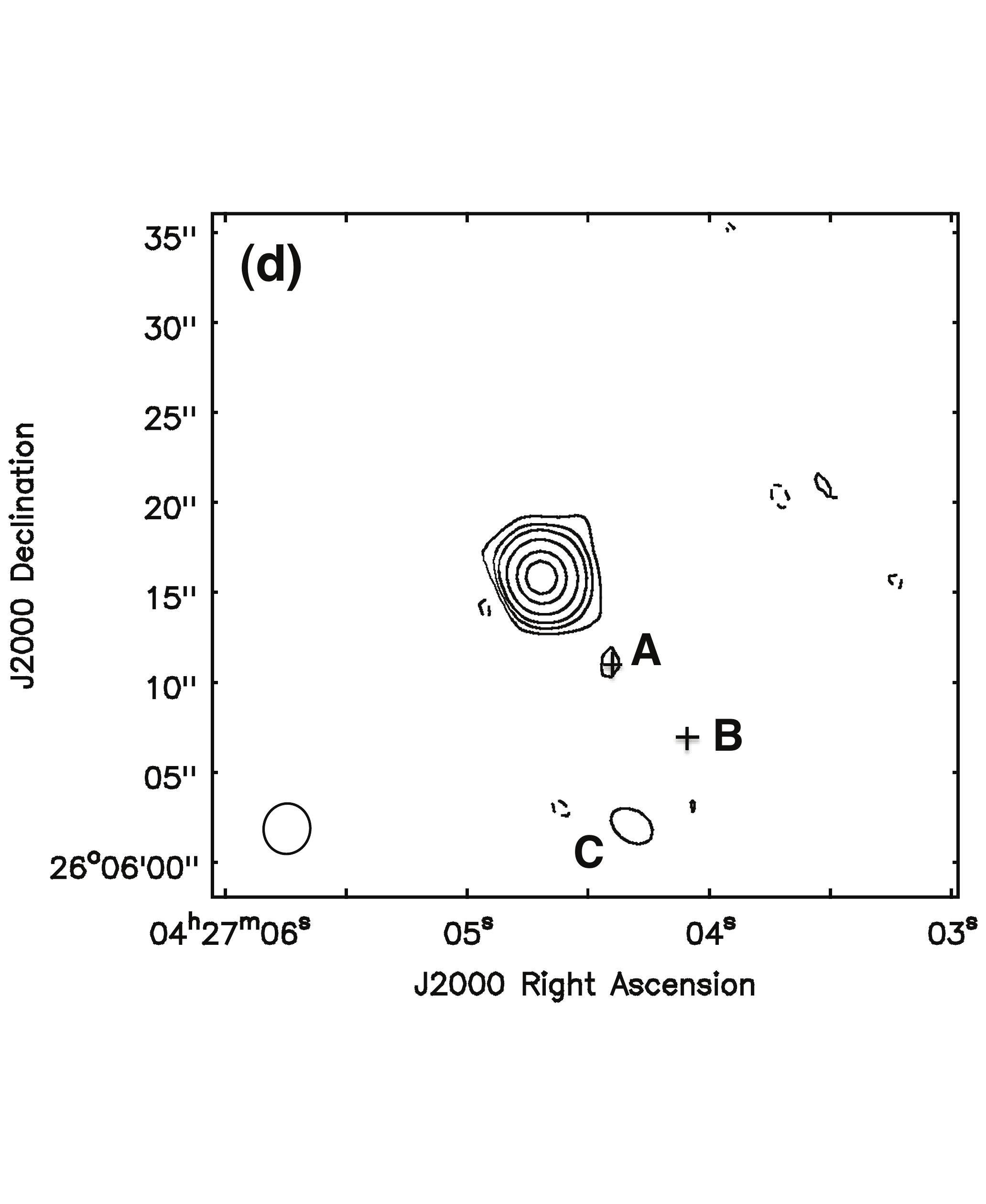}
\end{array}$
\end{center}
\caption{DG Tau contour maps at 8~GHz (a, c) and 5~GHz (b, d) at epochs 2012.29 and 2012.22 respectively.  The two maps in the first column of this figure, (a) \& (b), are made using natural weighting. The contour levels are -3, 3, 4, 6, 10, 20, 40, 60, 80, 100, 120, and 140 times the RMS in each map; the RMS and beam dimensions are given in Table (\ref{table:obs}). The second column maps, (c) \& (d), are made using uniform weighting in order to maximize angular resolution of the maps. The contour levels are -3, 3, 6, 10, 20, 40, and 60 times the RMS in each map. The RMS for (c) is 15 $\mu$Jy with beam dimensions $1.8\arcsec$ $\times$ $1.7\arcsec$; (d) has a RMS of 13 $\mu$Jy and beam dimensions of $2.9\arcsec$ $\times$ $2.6\arcsec$. The crosses in each of these maps indicate the locations of the two knots detected by \protect \citet{Rodriguez:2012}.} 
\label{fig:CXBand}
\end{figure}

Radio observations of DG Tau  at epoch 2009.6 \citep{Rodriguez:2012} show a radio knot located approximately 7\arcsec\ along the jet with an integrated flux density 150 $\mu$Jy, coincident with an optical [S II] knot. There is also a much weaker feature near 12\arcsec, also coincident with an optical knot. If these radio structures are associated with shock-compressed gas in the jet flow, we would expect them to evolve both in flux and position on the dynamical timescale of the shock and the expanding post-shock gas. 

Our observations confirm the existence of the inner knot. Figure \ref{fig:CXBand} shows 5.4~GHz  and 8.5~GHz images of DG Tau at epochs 2012.22 and 2012.29 respectively. At each frequency we show maps made with two different uv-plane density weighting functions: uniform weighting to maximize angular resolution (panels c, d), and natural (a.k.a. unity) weighting to maximize sensitivity to low-brightness features (panels a, b). 
The locations of the 7\arcsec\ knot (labeled A) and 12\arcsec\ knot (labeled B) detected by \citet{Rodriguez:2012} are indicated by crosses. Both the 8.5 and 5.4 GHz natural weighted maps show radio emission coincident with knot A but not with knot B. We did not detect either knot A or B in our epoch 2011.46 A-array observation, probably because the knots were over-resolved, as discussed below.  

\subsubsection{Knot proper motion}
The DG Tau jet optical knots  move outward with sky-plane proper motions $\sim0.3$\arcsec per year ($V\sim200$ km/s), at least within 10 arcsec of the stellar position \citep{Eisloffel:1998,Dougados:2000}. \citet{Rodriguez:2012} found radio knots that are approximately cospatial with the optical knots, and also move with this speed, suggesting that the radio and optical knots are two manifestations of the same traveling shocks. Here we examine whether our more recent knot observations are consistent with this hypothesis.  We consider only the motion of knot A, since we did not detect knot B at any epoch. Note that all positions discussed below are angular separations from the stellar position in the sky plane.
\begin{figure}[!ht]
\centerline{\includegraphics[width = 6in]{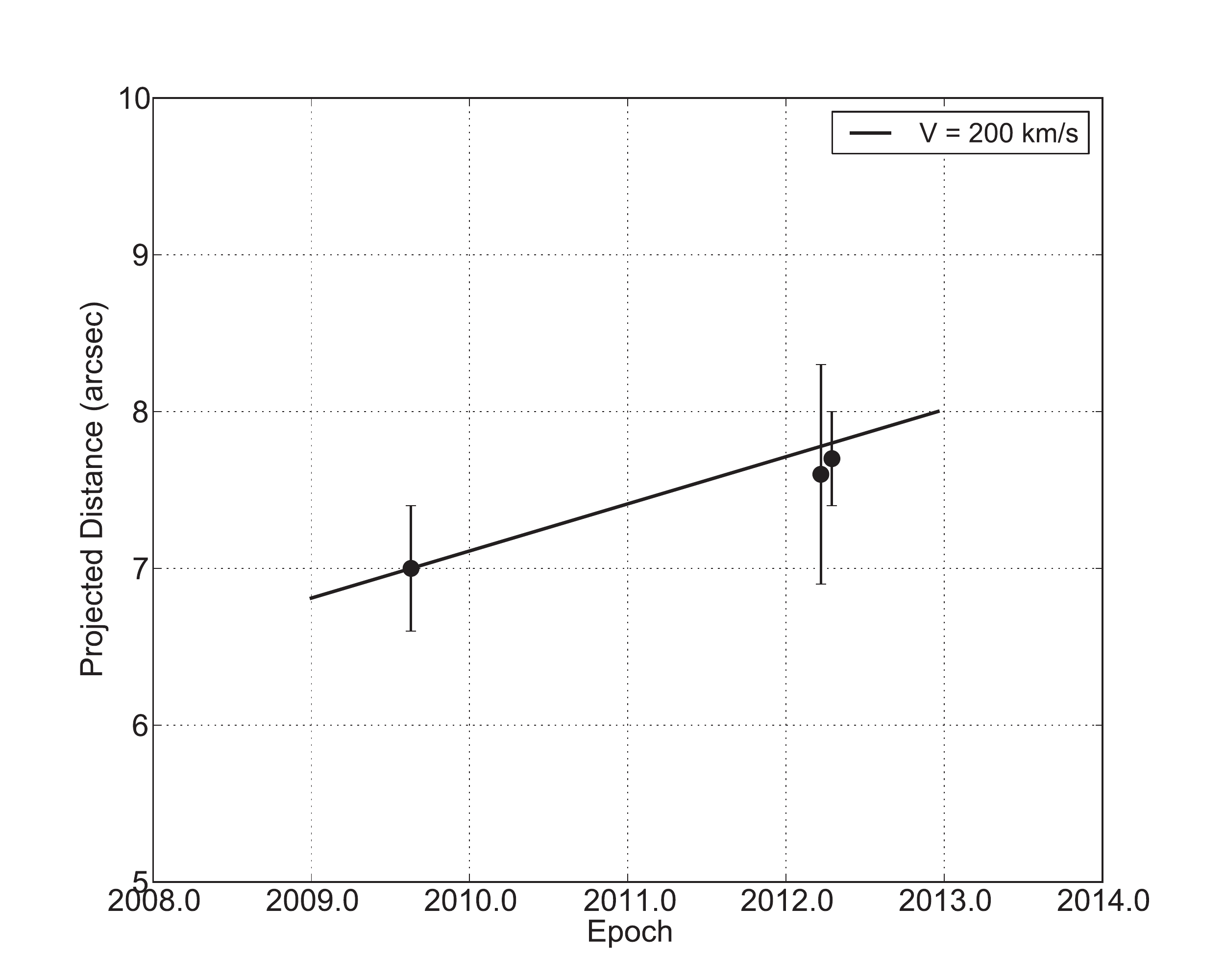}}
\caption{Projected knot location along jet vs. epoch (solid line) with observed centroid locations at epochs 2009.63 and 2012.22}
\label{fig:position-vs-epoch}
\end{figure}
\clearpage
Using the \protect\citet{Rodriguez:2012} calculated speed $V=198$ km s$^{-1}$ and their 2009.62 position for knot A, the expected position at epochs 2011.22 and 2011.29 is $7.8\arcsec$. We fit single-component GAUssian models to knot A in the 5.5~GHz  image. At 5.5~GHz, knot A has a centroid position $7.6\arcsec$$\ \pm\ $ $0.7\arcsec$, while at 8.5~GHz  it is $7.7\arcsec$$\ \pm\ $$0.3\arcsec$ (Figure~\ref{fig:position-vs-epoch}). Both positions agree with the predicted proper motion, but the fractional uncertainty is large, 37\% in the higher resolution 8.5~GHz image.

The optical position of knot A has recently been determined from HST STIS observations at epoch 2011.13 (Schneider et al. 2012, in preparation). Fitting a Gaussian to the spatial profile of the [S II] 6731$\AA$ line within $v$= -265 $\pm$ 65 km/s results in a distance $7.2\arcsec$ $\pm$ $0.1\arcsec$, compared with a predicted position $7.4\arcsec$ using the \citet{Rodriguez:2012} proper motion. The measured position is only slightly offset from the predicted position, and could indicate that the peak radio and optical intensities occur along different lines of sight in the shocked emission region. This is plausible, since the radio and optical line emergent intensities have different dependencies on the density and temperature of the gas.

\subsubsection{Knot evolution: Expanding post-shock gas?}

In addition to moving with the predicted proper motion, the flux density of knot A decreased dramatically.  In order to avoid differences associated with spectral index and angular resolution, we can compare the epoch 2009.63 measurement of \citet{Rodriguez:2012} with our 2012.29 observation, both of which were at 8.5~GHz and had the same angular resolution (VLA at C-array). The epoch 2009.63 integrated flux was $(150\ \pm \ 20)\ \mu$Jy while the 2012.29 integrated flux was $(46\ \pm \ 26)\ \mu$Jy, both measured using naturally-weighted maps. This is a 70\% decrease in 2.7 years, and is certainly real given the matched frequency and angular resolutions of the two observations. This behavior is not unprecedented: Radio knots in several other YSO jets have been observed to decrease with time and become undetectable within years of their initial ejection \citep[e.g.][]{Marti:1998}. The total flux density at 5.5~GHz (epoch 2012.22) was $(73\ \pm \ 25)\ \mu$Jy. This implies a spectral index $-0.4\ \pm\ 0.4$, consistent with optically thin thermal emission, but highly uncertain because of the large uncertainty in each measurement, and the differing angular resolutions. 

\citet{Rodriguez:2012} proposed a model consisting of periodic generation of shocks moving outward at at projected speed 200 km s$^{-1}$ in a conical outflow. The model predicts a periodic variation of knot flux density caused by corresponding periodic velocity variations in a bipolar outflow. 
Fig.~8 of \citet{Rodriguez:2012} predicts a flux for knot A near that observed at epoch 2012.29, but occurring 1.5 years after the flux level observed at epoch 2009.63 rather than 2.7 years. However, since their model is parameterized by several geometrical parameters which are not well-constrained, a detailed comparison is not appropriate with only a single additional flux measurement. Instead, guided by Occam's razor, we interpret the flux decrease using a much simpler conceptual model of an expanding volume of post-shock gas characterized by a single variable, the electron density. 

\begin{figure}
\centerline{\includegraphics[width=7.5in]{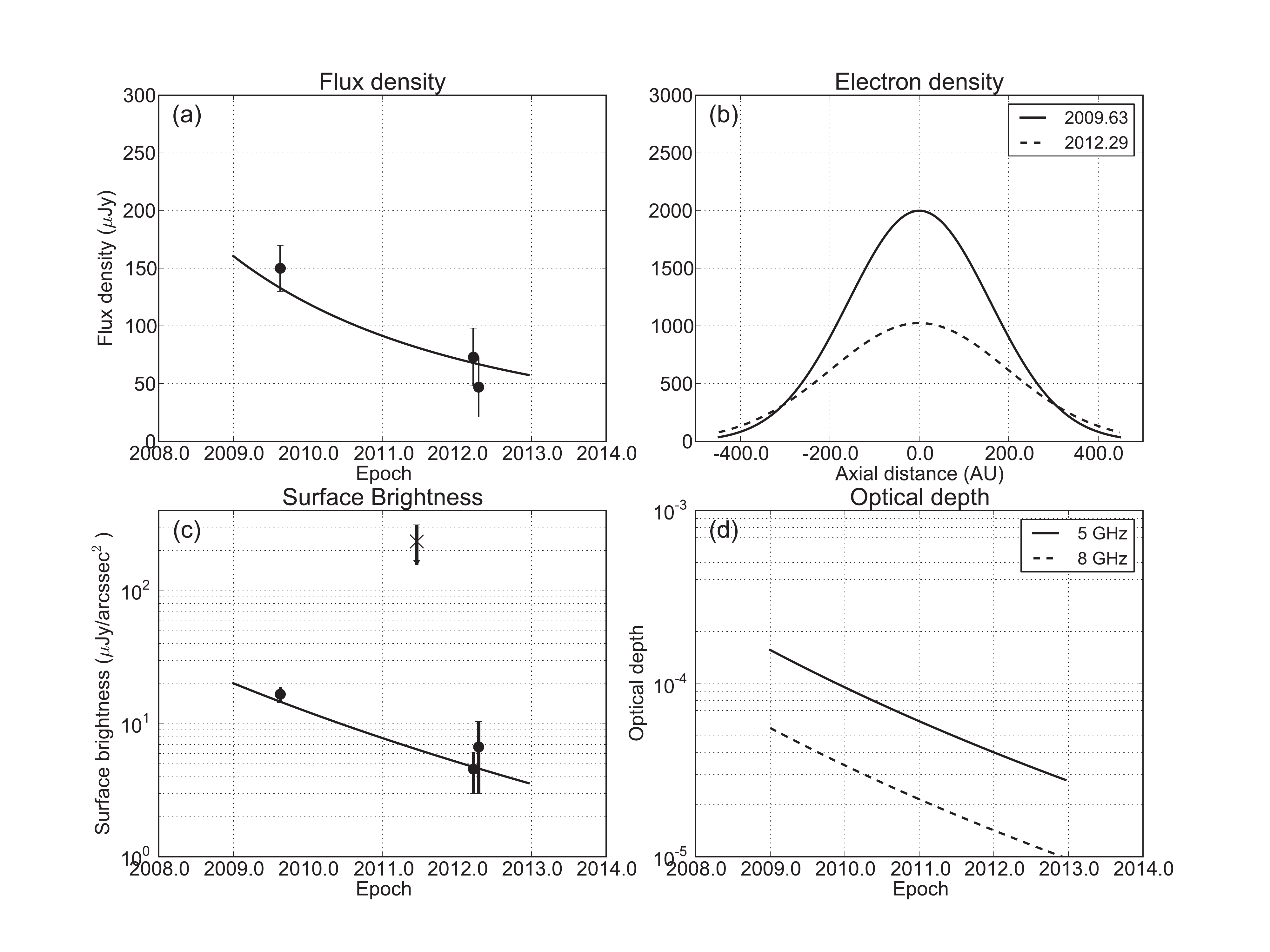}}
\caption{(a) DG Tau knot A flux density vs. epoch calculated using optically thin expanding sphere model (see text), with observed values at epochs 2009.63 \protect \citep{Rodriguez:2012} and 2012.22 (this paper). (b) Model electron density vs. projected axial distance (arc sec) at epochs 2009.63 (solid line) and 2012.22 (dashed line). (c) Model surface brightness vs. epoch (solid line) with observed values ($\bullet$) and upper limits ($\times$). (d) Optical depth at 5~GHz (solid line) and 8~GHz (dashed line) vs. epoch, confirming the optically-thin model assumption.}
\label{fig:model}
\end{figure}

We model the knot as an optically-thin, free-free emitting isotropic sphere whose density scales exponentially with radial distance $\rho$,
\begin{equation}
n_e(\rho)=\frac{N}  { {\rho_F}^{3 }  \sqrt{3{\pi}^3}   }    \ exp   \left[    -ln(2)  {  \left( \frac  {\rho}  {\rho_{F}  }   \right)^2}     \right],
 \end{equation}
 where $2\rho_F$ is the full width at half maximum scale, and the normalization ensures that the integrated density is a constant $N$, i.e., the total number of electrons does not change as the sphere expands.  We also assume the knot is isothermal, a simplistic but reasonable assumption since for optically-thin emission, the flux density dependence on temperature is much smaller than for density ($S\propto T^{-0.35}{n_e}^2$). We assume $T=10^4$ K, based on optical line ratio observations \citep{Agra-Amboage:2011}. 
 The sphere is assumed to expand linearly with time, $\rho(t) = \rho_0 + \dot{\rho} (t -t_0)$. 

The flux and size of the sphere were calculated as a function of time and compared with observations.  We solved for best-fit model parameters $N = 4.2\times10^{50}$, $\rho_0= 336$ AU($2.4\arcsec$ at d = 140 pc), and $\dot{\rho}$ = 21 AU yr$^{-1}$ (0.15$\arcsec\ yr^{-1}$). 
Figure~\ref{fig:model} shows the flux density, electron density profile, surface brightness, and optical depth as a function of epoch for the fitted model, along with observed values and upper limits. The agreement is well within the measurement uncertainty of the radio knot at all epochs, and has $\tau\ll 1$, as expected. 

\citet{Maurri:2012} measure electron densities $n_e\sim10^3$ cm$^{-3}$ along the jet at a projected distance near 4-5\arcsec. For a conical flow with constant ionization fraction, the density scales as $r^{-2}$, so we expect $n_e\sim500$ cm$^{-3}$ outside the radio knot at a projected distance of 7\arcsec. According to the model, the  maximum knot density varied from 2000 to 1000  cm$^{-3}$ from epochs 2009.6 -- 2012.2, implying  a density enhancement factor of 4 in 2009.6, decreasing to 2 in 2012.2. Assuming that the present expansion continues, the model predicts that the density contrast will vanish and that the knot will disappear within a few years.

\section{Summary}
We report multi-epoch VLA observations of the pre-main sequence star DG Tau's radio jet. The radio spectrum ($\alpha =0.46\ \pm\ $0.05) and lack of polarization indicate that the emission is bremsstrahlung, with no evidence for a non-thermal coronal component. Assuming the end of the radio jet at 0.35\arcsec\ (50 AU projected distance) is the $\tau =1$ surface, we calculate the column emission measure. Assuming an 'onion skin' ionization model and azimuthal symmetry, we find that the mean electron density in the centre of the jet is \={n} = 2.5$\times10^6$ cm$^{-3}$. This agrees well with optical estimates at this location \citep{Maurri:2012}. We model the electron density and velocity axial profiles to calculate the mass loss of the ionized component, $ {\dot{M}}_{ion}\sim$ 5 $\times$\ $10^{-8}$\ $M_\odot$\ $yr^{-1}$, with more than half of the mass flux in the high-velocity component within 5 AU of the jet center. This mass loss is  comparable to the total mass loss calculated using optical line observations \citep[e.g.,][]{Agra-Amboage:2011,Maurri:2012}, indicating that most of the mass loss in the jet at this location is in the ionized component.  

We confirm the existence of a radio knot near 7\arcsec\ recently reported by \citet{Rodriguez:2012}. The knot proper motion is consistent with a projected speed $V=200$ km $s^{-1}$, as suggested by \citet{Rodriguez:2012} and previous optical estimates. The flux density of the knot dramatically decreased between 2009.6 and 2012.2. We present a simple model for radio emission from the knot consisting of an optically-thin ionized sphere which expands linearly with time. The model predicts that the FWHM size of the radio knot increases from 340 AU (2.4\arcsec) to 390 AU (2.75\arcsec) from 2009.6 to 2012.2, while the central electron density decreases from 2000 to 1000 cm$^{-3}$. The resulting radio flux decreases from 150 $\mu$Jy to $50$ $\mu$Jy, in agreement with observations. By scaling from previously published density measurements at closer distances along the jet, we find that the density enhancement factor of the knot decreased from 4 in 2009.6  to 2 in 2012.2, and that the knot will disappear completely within a few years.

\bibliography{yso-rlm}

\begin{thebibliography}{51}
\expandafter\ifx\csname natexlab\endcsname\relax\def\natexlab#1{#1}\fi

\bibitem[{{Agra-Amboage} {et~al.}(2011){Agra-Amboage}, {Dougados}, {Cabrit}, \&
  {Reunanen}}]{Agra-Amboage:2011}
{Agra-Amboage}, V., {Dougados}, C., {Cabrit}, S., \& {Reunanen}, J. 2011, \aap,
  532, A59

\bibitem[{Andre(1996)}]{Andre:1996}
Andre, P. 1996, ASP Conference Series, 93, 273

\bibitem[{{Andre} {et~al.}(1992){Andre}, {Deeney}, {Phillips}, \&
  {Lestrade}}]{Andre:1992}
{Andre}, P., {Deeney}, B.~D., {Phillips}, R.~B., \& {Lestrade}, J.-F. 1992,
  \apj, 401, 667

\bibitem[{Andre {et~al.}(1988)Andre, Montmerle, {Feigelson}, Stine, \&
  Klein}]{Andre:1988}
Andre, P., Montmerle, T., {Feigelson}, E.~D., Stine, P.~C., \& Klein, K. 1988,
  \apj, 335, 940

\bibitem[{{Bacciotti} {et~al.}(1999){Bacciotti}, {Eisl{\"o}ffel}, \&
  {Ray}}]{Bacciotti:1999}
{Bacciotti}, F., {Eisl{\"o}ffel}, J., \& {Ray}, T.~P. 1999, \aap, 350, 917

\bibitem[{Bacciotti {et~al.}(2000)Bacciotti, Mundt, Ray, Eisl{\"o}ffel, Solf,
  \& Camezind}]{Bacciotti:2000}
Bacciotti, F., Mundt, R., Ray, T.~P., Eisl{\"o}ffel, J., Solf, J., \& Camezind,
  M. 2000, \apj, 537, L49

\bibitem[{{Bacciotti} {et~al.}(2002){Bacciotti}, {Ray}, {Mundt},
  {Eisl{\"o}ffel}, \& {Solf}}]{Bacciotti:2002b}
{Bacciotti}, F., {Ray}, T.~P., {Mundt}, R., {Eisl{\"o}ffel}, J., \& {Solf}, J.
  2002, \apj, 576, 222

\bibitem[{Bacciotti {et~al.}(2002)Bacciotti, Ray, Mundt, Eisl{\"o}ffel, \&
  Solf}]{Bacciotti:2002}
Bacciotti, F., Ray, T.~P., Mundt, R., Eisl{\"o}ffel, J., \& Solf, J. 2002,
  \apj, 576, 222

\bibitem[{{Cai} {et~al.}(2008){Cai}, {Shang}, {Lin}, \& {Shu}}]{Cai:2008}
{Cai}, M.~J., {Shang}, H., {Lin}, H.-H., \& {Shu}, F.~H. 2008, \apj, 672, 489

\bibitem[{{Carrasco-Gonzalez} {et~al.}(2010){Carrasco-Gonzalez}, Rodriguez,
  Anglada, Marti, Torrelles, \& Osorio}]{Carrasco:2010}
{Carrasco-Gonzalez}, C., Rodriguez, L.~F., Anglada, G., Marti, J., Torrelles,
  J.~M., \& Osorio, M. 2010, Science, 330, 1209

\bibitem[{{Choi} {et~al.}(2008){Choi}, {Hamaguchi}, {Lee}, \&
  {Tatematsu}}]{Choi:2008}
{Choi}, M., {Hamaguchi}, K., {Lee}, J.-E., \& {Tatematsu}, K. 2008, \apj, 687,
  406

\bibitem[{{Coffey} {et~al.}(2008){Coffey}, {Bacciotti}, \&
  {Podio}}]{Coffey:2008}
{Coffey}, D., {Bacciotti}, F., \& {Podio}, L. 2008, \apj, 689, 1112

\bibitem[{Cohen \& Bieging(1986)}]{Cohen:1986}
Cohen, M., \& Bieging, J. 1986, \aj, 92, 1396

\bibitem[{{Cohen} {et~al.}(1982){Cohen}, {Bieging}, \& {Schwartz}}]{Cohen:1982}
{Cohen}, M., {Bieging}, J.~H., \& {Schwartz}, P.~R. 1982, \apj, 253, 707

\bibitem[{Curiel {et~al.}(1993)Curiel, {Rodriguez}, Moran, \&
  {Canto}}]{Curiel:1993}
Curiel, S., {Rodriguez}, L.~F., Moran, J.~M., \& {Canto}, J. 1993, \apj, 415,
  191

\bibitem[{{Dougados} {et~al.}(2000){Dougados}, {Cabrit}, {Lavalley}, \&
  {M{\'e}nard}}]{Dougados:2000}
{Dougados}, C., {Cabrit}, S., {Lavalley}, C., \& {M{\'e}nard}, F. 2000, \aap,
  357, L61

\bibitem[{{Eisl{\"o}ffel} \& {Mundt}(1998)}]{Eisloffel:1998}
{Eisl{\"o}ffel}, J., \& {Mundt}, R. 1998, \aj, 115, 1554

\bibitem[{{Eisl{\"o}ffel} {et~al.}(2000){Eisl{\"o}ffel}, {Mundt}, {Ray}, \&
  {Rodriguez}}]{Eisloffel:2000}
{Eisl{\"o}ffel}, J., {Mundt}, R., {Ray}, T.~P., \& {Rodriguez}, L.~F. 2000,
  Protostars and Planets IV, 815

\bibitem[{{Feigelson} {et~al.}(1994){Feigelson}, Welty, Imhoff, Hall, Etzel, \&
  Phillips}]{Feigelson:1994}
{Feigelson}, E.~D., Welty, A.~D., Imhoff, C.~L., Hall, J.~C., Etzel, P.~B., \&
  Phillips, R. B.and~Londsdale, C.~J. 1994, \apj, 432, 373

\bibitem[{G{\"u}del {et~al.}(2005)G{\"u}del, Skinner, Briggs, Audard, Arzner,
  \& Telleschi}]{Gudel:2005}
G{\"u}del, M., Skinner, S.~L., Briggs, K.~R., Audard, M., Arzner, K., \&
  Telleschi, A. 2005, \apj, 626, L53

\bibitem[{{G{\"u}del} {et~al.}(2007){G{\"u}del}, {Telleschi}, {Audard},
  {Skinner}, {Briggs}, {Palla}, \& {Dougados}}]{Gudel:2007}
{G{\"u}del}, M., {Telleschi}, A., {Audard}, M., {Skinner}, S.~L., {Briggs},
  K.~R., {Palla}, F., \& {Dougados}, C. 2007, \aap, 468, 515

\bibitem[{Hartigan {et~al.}(1995)Hartigan, Edwards, \&
  Ghandour}]{Hartigan:1995}
Hartigan, P., Edwards, S., \& Ghandour, L. 1995, The Astrophysical Journal,
  452, 736

\bibitem[{Hughes(1997)}]{Hughes:1997}
Hughes, V.~A. 1997, \apj, 481, 857

\bibitem[{{Kitamura} {et~al.}(1996{\natexlab{a}}){Kitamura}, {Kawabe}, \&
  {Saito}}]{Kitamura:1996a}
{Kitamura}, Y., {Kawabe}, R., \& {Saito}, M. 1996{\natexlab{a}}, \apj, 457, 277

\bibitem[{{Kitamura} {et~al.}(1996{\natexlab{b}}){Kitamura}, {Kawabe}, \&
  {Saito}}]{Kitamura:1996}
---. 1996{\natexlab{b}}, \apjl, 465, L137

\bibitem[{Lavalley {et~al.}(1997)Lavalley, Cabrit, Dougados, Ferruit, \&
  Bacon}]{Lavalley:1997}
Lavalley, C., Cabrit, S., Dougados, C., Ferruit, P., \& Bacon, R. 1997, \aap,
  327, 671

\bibitem[{Lavalley-Fouquet {et~al.}(2000)Lavalley-Fouquet, Cabrit, \&
  Dougados}]{Lavalley-Fouquet:2000}
Lavalley-Fouquet, C., Cabrit, S., \& Dougados, C. 2000, \aap, 356, L41

\bibitem[{{Mart{\'{i}}} {et~al.}(1998){Mart{\'{i}}}, {Rodr{\'{i}}guez}, \&
  {Reipurth}}]{Marti:1998}
{Mart{\'{i}}}, J., {Rodr{\'{i}}guez}, L.~F., \& {Reipurth}, B. 1998, \apj, 502,
  337

\bibitem[{Maurri {et~al.}(2012)Maurri, Bacciotti, Podio, Eisoffel, Ray, Mundt,
  Locatelli, \& Coffey}]{Maurri:2012}
Maurri, L., Bacciotti, F., Podio, L., Eisoffel, J., Ray, T., Mundt, R.,
  Locatelli, U., \& Coffey, D. 2012, Astron. Astrophys., submitted

\bibitem[{{McGroarty} \& {Ray}(2004)}]{McGroarty:2004}
{McGroarty}, F., \& {Ray}, T.~P. 2004, \aap, 420, 975

\bibitem[{{McGroarty} {et~al.}(2007){McGroarty}, {Ray}, \&
  {Froebrich}}]{McGroarty:2007}
{McGroarty}, F., {Ray}, T.~P., \& {Froebrich}, D. 2007, \aap, 467, 1197

\bibitem[{{Mundt} \& {Fried}(1983)}]{Mundt:1983}
{Mundt}, R., \& {Fried}, J.~W. 1983, \apjl, 274, L83

\bibitem[{Osten \& Wolk(2009)}]{Osten:2009}
Osten, R.~A., \& Wolk, S.~J. 2009, \apj, 691, 1128

\bibitem[{Phillips {et~al.}(1993)Phillips, Lonsdale, \&
  {Feigelson}}]{Phillips:1993}
Phillips, R.~B., Lonsdale, C.~J., \& {Feigelson}, E.~D. 1993, \apj, 403, L43

\bibitem[{Phillips {et~al.}(1996)Phillips, Lonsdale, {Feigelson}, \&
  Deeney}]{Phillips:1996}
Phillips, R.~B., Lonsdale, C.~J., {Feigelson}, E.~D., \& Deeney, B.~D. 1996,
  \aj, 111, 918

\bibitem[{{Pudritz} {et~al.}(2012){Pudritz}, {Hardcastle}, \&
  {Gabuzda}}]{Pudritz:2012}
{Pudritz}, R.~E., {Hardcastle}, M.~J., \& {Gabuzda}, D.~C. 2012, \ssr, 169, 27

\bibitem[{{Pyo} {et~al.}(2003){Pyo}, {Kobayashi}, {Hayashi}, {Terada}, {Goto},
  {Takami}, {Takato}, {Gaessler}, {Usuda}, {Yamashita}, {Tokunaga}, {Hayano},
  {Kamata}, {Iye}, \& {Minowa}}]{Pyo:2003a}
{Pyo}, T.-S., {Kobayashi}, N., {Hayashi}, M., {Terada}, H., {Goto}, M.,
  {Takami}, H., {Takato}, N., {Gaessler}, W., {Usuda}, T., {Yamashita}, T.,
  {Tokunaga}, A.~T., {Hayano}, Y., {Kamata}, Y., {Iye}, M., \& {Minowa}, Y.
  2003, \apj, 590, 340

\bibitem[{{Ray} {et~al.}(1997){Ray}, {Muxlow}, {Axon}, {Brown}, {Corcoran},
  {Dyson}, \& {Mundt}}]{Ray:1997}
{Ray}, T.~P., {Muxlow}, T.~W.~B., {Axon}, D.~J., {Brown}, A., {Corcoran}, D.,
  {Dyson}, J., \& {Mundt}, R. 1997, \nat, 385, 415

\bibitem[{{Reynolds}(1986)}]{Reynolds:1986}
{Reynolds}, S.~P. 1986, \apj, 304, 713

\bibitem[{{Rodr{\'{\i}}guez} {et~al.}(1999){Rodr{\'{\i}}guez}, Anglada, \&
  Curiel}]{Rodriguez:1999}
{Rodr{\'{\i}}guez}, L.~F., Anglada, G., \& Curiel, S. 1999, \apj, 125, 427

\bibitem[{{Rodr{\'{\i}}guez} {et~al.}(2005){Rodr{\'{\i}}guez}, Garay, Brooks,
  \& Mardones}]{Rodriguez:2005}
{Rodr{\'{\i}}guez}, L.~F., Garay, G., Brooks, K.~J., \& Mardones, D. 2005,
  \apj, 629, 953

\bibitem[{{Rodr{\'{\i}}guez} {et~al.}(2012){Rodr{\'{\i}}guez}, {Gonz{\'a}lez},
  {Raga}, {Cant{\'o}}, {Riera}, {Loinard}, {Dzib}, \&
  {Zapata}}]{Rodriguez:2012}
{Rodr{\'{\i}}guez}, L.~F., {Gonz{\'a}lez}, R.~F., {Raga}, A.~C., {Cant{\'o}},
  J., {Riera}, A., {Loinard}, L., {Dzib}, S.~A., \& {Zapata}, L.~A. 2012, \aap,
  537, A123

\bibitem[{{Schneider} {et~al.}(2011){Schneider}, {G{\"u}nther}, \&
  {Schmitt}}]{Schneider:2011}
{Schneider}, P.~C., {G{\"u}nther}, H.~M., \& {Schmitt}, J.~H.~M.~M. 2011, \aap,
  530, A123

\bibitem[{{Schneider} \& {Schmitt}(2008)}]{Schneider:2008}
{Schneider}, P.~C., \& {Schmitt}, J.~H.~M.~M. 2008, \aap, 488, L13

\bibitem[{Skinner(1993)}]{Skinner:1993}
Skinner, S.~L. 1993, \apj, 408

\bibitem[{Takami {et~al.}(2004)Takami, Chrysostomou, Ray, Davis, Dent, Bailey,
  Tamura, \& Terada}]{Takami:2004}
Takami, M., Chrysostomou, A., Ray, T.~P., Davis, C., Dent, W. R.~F., Bailey,
  J., Tamura, M., \& Terada, H. 2004, A\&A, 416, 213

\bibitem[{{Torres} {et~al.}(2009){Torres}, {Loinard}, {Mioduszewski}, \&
  {Rodr{\'{\i}}guez}}]{Torres:2009}
{Torres}, R.~M., {Loinard}, L., {Mioduszewski}, A.~J., \& {Rodr{\'{\i}}guez},
  L.~F. 2009, \apj, 698, 242

\bibitem[{White {et~al.}(1992)White, Pallavicini, \& Kundu}]{White:1992}
White, S.~M., Pallavicini, R., \& Kundu, M.~R. 1992, \aap, 259, 149

\bibitem[{Wilner {et~al.}(1999)Wilner, Reid, \& Menten}]{Wilner:1999}
Wilner, D.~J., Reid, M.~J., \& Menten, K.~M. 1999, \apj, 513, 775

\bibitem[{{Yusef-Zadeh} {et~al.}(1990){Yusef-Zadeh}, {Cornwell}, {Reipurth}, \&
  {Roth}}]{Yusef-Zadeh:1990}
{Yusef-Zadeh}, F., {Cornwell}, T.~J., {Reipurth}, B., \& {Roth}, M. 1990,
  \apjl, 348, L61

\bibitem[{{Zapata} {et~al.}(2004){Zapata}, {Rodriguez}, \& Kurtz}]{Zapata:2004}
{Zapata}, L.~A., {Rodriguez}, L.~F., \& Kurtz, S.~E. 2004, \apj, 127, 2252

\end{thebibliography}

\end{document}